\documentclass{article}
\usepackage{spconf,amssymb,amsmath,graphicx,booktabs}
\usepackage{color}
\usepackage{multirow}
\usepackage{makecell}
\usepackage{url}
\usepackage[hidelinks]{hyperref}
\usepackage{subcaption}
\usepackage{caption} 
\usepackage[utf8]{inputenc} %

\ninept
\makeatletter
\def\@IEEEsectpunct{.\ \,}
\def\paragraph{\@startsection{paragraph}{4}{\z@}{1.5ex plus 0.5ex minus 0.2ex}%
{-1em}{\normalfont\normalsize\bf}}
\makeatother

\renewenvironment{thebibliography}[1]{
  \begin{oldthebibliography}{#1}
  \setlength{\itemsep}{0.1em}
  \setlength{\parskip}{0.1em}
}
{
  \end{oldthebibliography}
}

\usepackage[edges]{forest}
\definecolor{folderbg}{RGB}{124,166,198}
\definecolor{folderborder}{RGB}{110,144,169}
\newlength\Size
\tikzset{%
  folder/.pic={%
    \filldraw [draw=folderborder, top color=folderbg!50, bottom color=folderbg] (-1.05*\Size,0.2\Size+3.75pt) rectangle ++(.75*\Size,-0.2\Size-3.75pt);
    \filldraw [draw=folderborder, top color=folderbg!50, bottom color=folderbg] (-1.15*\Size,-\Size) rectangle (1.15*\Size,\Size);
  },
  file/.pic={%
    \filldraw [draw=folderborder, top color=folderbg!5, bottom color=folderbg!10] (-\Size,.4*\Size+3.75pt) coordinate (a) |- (\Size,-1.2*\Size) coordinate (b) -- ++(0,1.6*\Size) coordinate (c) -- ++(-3.75pt,3.75pt) coordinate (d) -- cycle (d) |- (c) ;
  },
}
\forestset{%
  declare autowrapped toks={pic me}{},
  pic dir tree/.style={%
    for tree={%
      folder,
      font=\ttfamily,
      grow'=0,
    },
    before typesetting nodes={%
      for tree={%
        edge label+/.option={pic me},
      },
    },
  },
  pic me set/.code n args=2{%
    \forestset{%
      #1/.style={%
        inner xsep=2\Size,
        pic me={pic {#2}},
      }
    }
  },
  pic me set={directory}{folder},
  pic me set={file}{file},
}

\title{ESPNET-SE: END-TO-END SPEECH ENHANCEMENT AND SEPARATION TOOLKIT DESIGNED FOR ASR INTEGRATION}
\name{\it Chenda Li$^{1*}$\thanks{$^*$These authors contributed equally to this work.}, Jing Shi$^{2,3*}$, Wangyou Zhang$^{1*}$, Aswin Shanmugam Subramanian$^3$, Xuankai Chang$^3$, \\
\it Naoyuki Kamo, Moto Hira$^4$, Tomoki Hayashi$^{5,6}$, Christoph Boeddeker$^7$, Zhuo Chen$^8$,  Shinji Watanabe$^3$}
\address{
    $^1$Shanghai Jiao Tong University,
    $^2$Institute of Automation, Chinese Academy of Sciences,\\
    $^3$Johns Hopkins University,
    $^4$Facebook AI,
    $^5$Nagoya University,\\
    $^6$Human Dataware Lab. Co., Ltd.,
    $^7$Paderborn University,
    $^8$Microsoft Research
}
\begin{document}
\ninept
\maketitle
\begin{abstract}
We present ESPnet-SE, which is designed for the quick development of speech enhancement and speech separation systems in a single framework, along with the optional downstream speech recognition module. ESPnet-SE is a new project which integrates rich automatic speech recognition related models, resources and systems to support and validate the proposed front-end implementation (i.e. speech enhancement and separation).
It is capable of processing both single-channel and multi-channel data, with various functionalities including dereverberation, denoising and source separation.
We provide all-in-one recipes including data pre-processing, feature extraction, training and evaluation pipelines for a wide range of benchmark datasets. This paper describes the design of the toolkit, several important functionalities, especially the speech recognition integration, which differentiates ESPnet-SE from other open source toolkits, and experimental results with major benchmark datasets. %

\end{abstract}
\section{Introduction}
\label{sec:intro}
As the core parts of the speech processing front-end, speech enhancement and separation (SE) have been studied for decades.
Speech enhancement tries to improve the intelligibility and quality of speech contaminated by additive noise and reverberation~\cite{loizou2013speech}, while speech separation focuses on the speech in multi-speaker conditions, which is also known as the ``cocktail party'' problem~\cite{cherry1953some}. Speech enhancement and separation are highly desirable for a vast range of applications, such as smart speakers, automatic meeting transcription, automatic captioning for audio/video recordings (e.g., YouTube), multi-party human-machine interaction (e.g., in the world of Internet of things (IoT)), and advanced hearing aids, where noisy speech is commonly encountered~\cite{gustafsson2001spectral,vincent2018audio,haeb2019speech,Kolbaek2017Multitalker}.

Recently with the great success of deep learning techniques, end-to-end speech enhancement and separation systems (hereinafter collectively referred to as E2E-SE) have grown in popularity and have even started replacing conventional systems described in~\cite{loizou2013speech} and ~\cite{makino2007blind}.
The E2E-SE system, taking the raw waveform of noisy speech as input, outputs one or several speech sources for each target speaker. This end-to-end paradigm can easily integrate signal processing algorithms along with various learnable neural networks, making full use of neural network toolkits.
Meanwhile, many downstream speech processing back-end tasks (i.e., automatic speech recognition, keyword spotting and speech translation)  can also benefit greatly by cascading after it.   

In this paper, we introduce a new E2E-SE toolkit named ESPnet-SE\footnote{The demo page is \url{https://colab.research.google.com/drive/1fjRJCh96SoYLZPRxsjF9VDv4Q2VoIckI}}, which is an extension of the open-source speech processing toolkit ESPnet~\cite{Espnet-Watanabe2018}. ESPnet-SE fully considers the various forms of speech input in the front-end scenes and meanwhile flexibly and organically integrated with the downstream automatic speech recognition (ASR) task, making it a user-friendly toolkit to easily build totally end-to-end robust ASR systems, even without need for clean speech signals. The toolkit provides adaptability to different speech data, including (1) single and multiple speakers, (2) single and multiple channels, (3) anechoic and reverberant conditions.
Moreover, thanks to the ripe and efficient ASR modules in ESPnet, rich speech recognition related models, resources and systems can be optionally concatenated after the E2E-SE system, enabling evaluation and joint optimization with ASR, which distinguishes ESPnet-SE from other open source toolkits. %
\section{Related Work}
\label{sec:relate}
\begin{table*}
\caption{Comparison with other open-source deep learning based speech enhancement and separation toolkits, where $\star$ denotes an on-going project in August 2020.}
\label{tab:related_work}
\centering
\begin{tabular}{l | c c c c c}
 \toprule
 & \href{https://github.com/nussl/nussl}{nussl} & \href{https://github.com/speechLabBcCuny/onssen}{Onssen} & \href{https://sigsep.github.io/open-unmix/}{Open-Unmix} & \href{https://github.com/mpariente/asteroid}{Asteroid} & \href{https://github.com/espnet/espnet}{ESPnet-SE} \\
 \midrule
 T-F masking & $\checkmark$ & $\checkmark$ & & $\checkmark$ & $\checkmark$ \\
 Chimera & $\checkmark$ & $\checkmark$ & & $\checkmark$ &  \\
 Deep clustering & $\checkmark$ & $\checkmark$ & & $\checkmark$ & $\star$ \\
 TasNet & $\star$ & $\checkmark$ & & $\checkmark$ & $\checkmark$ \\
 DPRNN & $\star$ & & & $\checkmark$ & $\checkmark$ \\
 Wavesplit & & & & $\star$ & $\star$ \\
 WPE &  & & & & $\checkmark$ \\
 Neural beamformer & & & & & $\checkmark$ \\
 OpenUnmix & $\star$ & & $\checkmark$ & $\star$ &  \\
 \midrule
 Provide data preparation? & & & $\checkmark$ & $\checkmark$ & $\checkmark$ \\
 Support training and evaluation? & $\checkmark$ & $\checkmark$ & $\checkmark$ & $\checkmark$ & $\checkmark$ \\
 Support speech source separation? & $\checkmark$ & $\checkmark$ & & $\checkmark$ & $\checkmark$ \\
 Support music source separation? & $\checkmark$ & & $\checkmark$ & $\star$ &  \\
 Support downloading pretrained model? & $\checkmark$ & $\star$ & $\checkmark$ & $\checkmark$ & $\checkmark$ \\
 Support integration with ASR? & & & & & $\checkmark$ \\
 \midrule
 \# of supported datasets & 2 & 3 & 5 & 10 & 10 \\
 \midrule
 Backend & PyTorch & PyTorch & PyTorch \& NNabla & PyTorch & PyTorch \\
 License & MIT & GPL-3.0 & MIT & MIT & Apache-2.0 \\
\bottomrule
\end{tabular}
\end{table*}

This section briefly compares ESPnet-SE to other open-source deep learning-based speech enhancement and separation toolkits. 
We pick up four well-maintained frameworks in this range, including \textit{nussl} (North-western University Source Separation Library) \cite{manilow2018northwestern}, \textit{Onssen} (An Open-source Speech Separation and Enhancement Library)~\cite{ni2019onssen}, \textit{Open-Unmix}~\cite{stoter2019open-unmix} and \textit{Asteroid} (Audio source separation on Steroids)~\cite{Asteroid-Pariente2020}. Other projects (e.g., \textit{padertorch}\footnote{\url{https://github.com/fgnt/padertorch}}) may also contain similar features in E2E-SE to some extent, but because they are not systematically designed to address these tasks or still under construction in the paper submission stage, no comparison is made here.

For the first two frameworks \textit{nussl} and \textit{Onssen}, they are based on the \textit{PyTorch}~\cite{PyTorch-Paszke2019} platform and provide several state-of-the-art SE methods, along with the training and evaluation scripts. However, data preparation steps are not provided in the recipes and experiments are not easily configurable from the command line~\cite{Asteroid-Pariente2020}.
On the other hand, \textit{Open-Unmix} and \textit{Asteroid} provide a whole pipeline from data preparation to evaluation with common line support. 
Note that the former is aimed for a specific model of \textit{Open-Unmix} on the music separation task. 
The same discussion is also applied to a series of isolated speech separation projects, such as \texttt{\href{https://github.com/kaituoxu/TasNet}{kaituoxu/TasNet}}, \texttt{\href{https://github.com/kaituoxu/TasNet}{kaituoxu/Conv-TasNet}}, \texttt{\href{https://github.com/yluo42/TAC}{yluo42/TAC}} and so on.
For \textit{Asteroid}, it is well-designed for the speech separation task, which supports a large range of datasets and architectures, and a set of recipes to reproduce some important papers. However, its overall framework still belongs to the category of the front-end, which outputs the separated or enhanced speech signals. 
Although this application itself is quite important, it would not be fully optimized for the downstream application, including ASR systems. Actually, with the gradual maturity of speech separation technologies, many works have paid their attention to the optimization of speech enhancement\,/\,separation using speech recognition objectives~\cite{haeb2019speech,Speech-Subramanian2019, MIMO-Chang2019, Location-Subramanian2020, chen2020continuous,von2020multi,CHiME4-Vincent2017,REVERB-Kinoshita2016}.

Taking into account this trend and the direction of future works in the community, we propose the ESPnet-SE toolkit with the design for ASR integration. In our framework, we provide the optional module of ASR to take part in the training and evaluation. Plenty of the well-established or customizable recognition models are provided based on both connectionist temporal classification (CTC) and attention-based encoder-decoder networks. Users can easily choose to use the recognition network for joint training with E2E-SE or to just apply the downloadable pretrained models to recognize the enhanced or separated speech signals. Even on some unsupervised or real collection of speech recordings (e.g. CHiME-4 ~\cite{CHiME4-Vincent2017}) without the reference speech signals, due to the existence of transcription, our systems can be used to train directly with the recognition loss.
The overall comparison is summarized in Table~\ref{tab:related_work}, and it can be seen that the functionalities of our toolkit are complementary to the other toolkits.

\section{Features of ESPnet-SE}
\label{sec:features}
The ESPnet-SE toolkit provides two main components: the implementation of different speech enhancement/separation models, and recipes for many common speech enhancement/separation datasets. The ESPnet-SE models are implemented using PyTorch \cite{PyTorch-Paszke2019} with rich features. The recipes are carefully designed to follow a unified pipeline, with a stage-by-stage processing style. Below we introduce each component in detail.

\subsection{Models}
\label{sec:models}
Our implemented ESPnet-SE models cover a wide range of scenarios, including (1) time-domain and frequency-domain; (2) single-channel and multi-channel; (3) single-source and multi-source.
For time domain models, we implemented the time-domain audio separation network (TasNet) \cite{Tasnet-Luo2018,Conv-Luo2019} and the dual-path RNN (DPRNN) \cite{luo2020dual} based variants.
For frequency-domain models, we support both time-frequency (T-F) masking \cite{Permutation-Yu2017} and neural beamformer \cite{Neural-Heymann2016,erdogan2016improved} networks.
Among these four models, TasNet and DPRNN operate on the single-channel data, while T-F masking and the neural beamformer can operate on both single-channel and multi-channel data.
All models support both speech enhancement for single-source input and speech separation for multi-source input.
The input for each model is the raw waveform, and the short-time Fourier transform (STFT) will be used in the frequency-domain models to transform the input to frequency domain.
All the processing including STFT is implemented with differentiable PyTorch functions. 

T-F masking is a widely used approach to speech enhancement and separation. It estimates a T-F mask for each source in the input, and the mask is used to reconstruct the STFT spectrum of each source. The T-F masking model consists of a recurrent neural network (RNN)\footnote{We can also easily replace it with other architectures from the ESPnet-ASR module, such as CNN, self-attention and conformer.} followed by several parallel linear layers, with each layer corresponding to one source.

TasNet enhances or separates the audio signal in the time domain. It uses a convolution encoder to encode the audio signal into high-level representations, and then perform enhancement or separation on it. The processed representations are then reconstructed back to audio signals by a deconvolution decoder. Recently, the dual-path RNN (DPRNN) based TasNet \cite{luo2020dual} achieves state-of-the-art performance on the WSJ0-2mix corpus \cite{Deep-Hershey2016}.

The neural beamformer is designed for multi-channel input. It consists of a dereverberation module (optional) and a beamforming module. Each module contains a T-F mask estimation network that is similar to the T-F masking model. The dereverberation module is based on the weighted prediction error (WPE) algorithm \cite{Speech-Yoshioka2010}, and the beamforming module supports various beamforming types, including minimum-variance distortionless response (MVDR) \cite{Optimal-Souden2009}, minimum-power distortionless response (MPDR) \cite{Optimum-Van2004}, and weighted power minimization distortionless response (WPD) \cite{Unified-Nakatani2019}.

\subsection{Training for SE and SS}
\label{sec:training_for_ss_se}
We support multiple loss functions for training the ESPnet-SE models, and they can be divided into three categories: mask approximation loss, signal approximation loss, and metric-based loss.

The mask approximation loss is mainly designed for models with mask estimation components, such as the T-F masking network and the neural beamformer. It measures the error between the estimated mask $\hat{m}$ and the target mask $m$. Both mean square error (MSE) and cross-entropy (CE) cost functions can be used.
We support various types of target masks, including ideal binary mask (IBM) \cite{IBM-Li2009}, ideal ratio mask (IRM) \cite{IRM-Narayanan2013}, ideal amplitude mask (IAM, also known as FFT-mask in \cite{OnTraining-Wang2014}) and phase-sensitive mask (PSM) \cite{PSM-Erdogan2015}.

The signal approximation loss measures the error between the enhanced signal $\hat{x}$ and the reference signal $x$. It can be calculated on either the magnitude spectrum or the complex spectrum. A typical cost function for signal approximation is the mean squared error.

The metric-based loss directly measures the quality of the enhanced signal with a specific metric, such as the scale-invariant signal-to-noise ratio (SI-SNR) \cite{Tasnet-Luo2018}.

Except for the time domain model that does not support the mask approximation loss, all models have implemented different \textit{forward} functions to support the aforementioned loss functions for training, and it can be easily configured through a configuration file.

For all types of loss functions in speech separation, the permutation invariant training (PIT) \cite{Permutation-Yu2017} is applied to solve the label ambiguity problem. In addition, we also support the unsupervised source separation method from the newly published work~\cite{wisdom2020unsupervised}.

\begin{figure}[t]
  \begin{minipage}{\linewidth}
    \centering
    \centerline{\includegraphics[width=\textwidth]{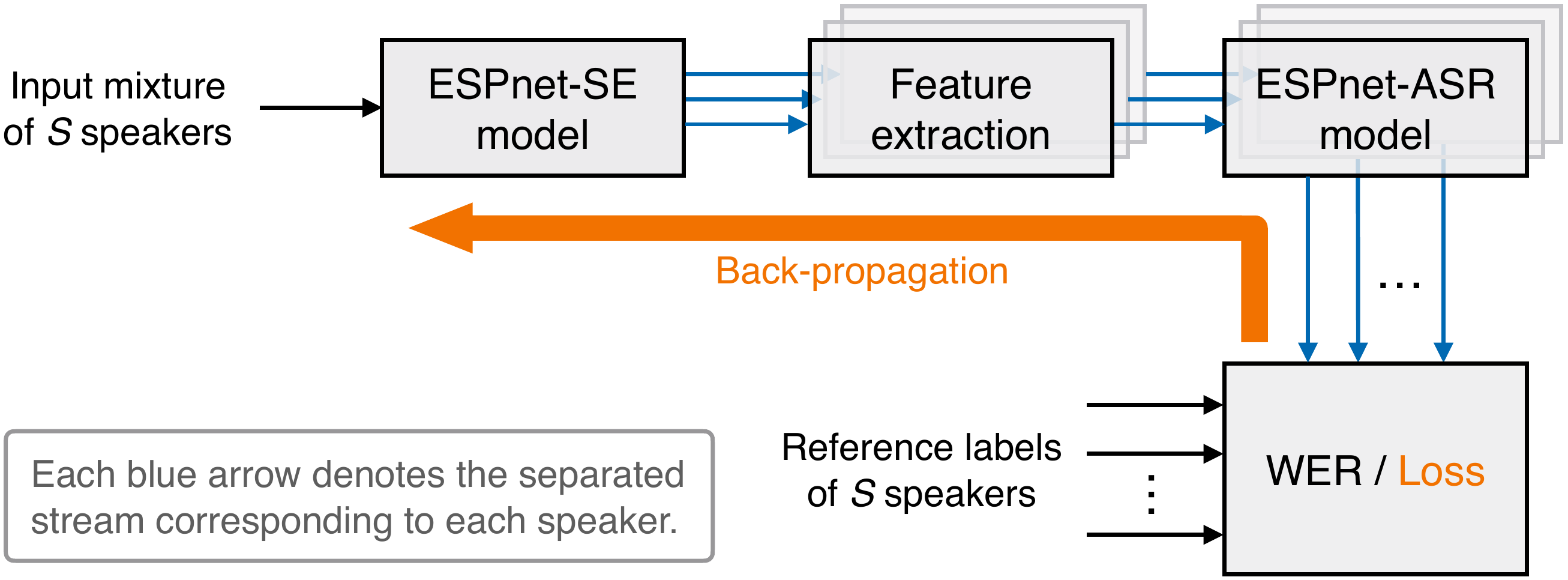}}
  \end{minipage}
\caption{Block diagram of ESPnet-SE models with ASR integration. The orange part shows the back-propagation process for joint optimization.}
\label{fig:integration_with_asr}
\end{figure}
\subsection{Evaluation for SE and SS}
\label{sec:evaluation_for_ss_se}

We evaluate the speech enhancement or separation with various commonly used evaluation metrics, which can be divided into two classes: signal-level metrics and perception-level metrics.
For signal-level measurement, the source-to-noise ratio (SNR), source-to-distortion ratio (SDR) \cite{vincent2006performance}, source-to-interference ratio (SIR) \cite{vincent2006performance} and source-to-artifact ratio (SAR) \cite{vincent2006performance} are used by default. The scale-invariant signal-to-noise ratio (SI-SNR) \cite{Tasnet-Luo2018} is also available.
For perception-level measurement, the default metric is the short-time objective intelligibility (STOI) \cite{taal2011algorithm} and optionally the perceptual evaluation of speech quality (PESQ) \cite{rix2001perceptual} is also available.

Furthermore, the enhanced or separated speech can be evaluated with pretrained single speaker ASR systems, which will be introduced in detail in the following subsection.

\subsection{Integration with ASR}
\label{sec:integration_with_asr}
Thanks to the unified design, we can easily enable the integration with ASR.

Firstly, the ESPnet-SE models can be directly evaluated with a pretrained ASR model, as illustrated in Fig.~\ref{fig:integration_with_asr}. The input mixture of $S$ speakers is first separated by the ESPnet-SE model into $S$ streams, corresponding to $S$ different speakers. Then we extract features (e.g.~FilterBank) for each stream separately, and feed them into the ASR model to generate the recognized token sequence. Finally, we calculate the word error rate (WER) of the recognition results. When $S > 1$, the final WER is determined with the best permutation between recognition results and reference labels, which minimizes the total WER of all speakers.

Furthermore, we can even jointly optimize both ESPnet-SE and the ASR models.
The forward process is the same as described above, after replacing WER with the ASR loss. And permutation invariant training is used when $S > 1$.
Since the forward operations are fully differentiable, the loss can be back-propagated from the ASR model to the ESPnet-SE model, so that the entire system can be jointly optimized.

\begin{figure}[t]
\begin{minipage}[b]{\linewidth}
\centering
\begin{forest}
  pic dir tree,
  where level=0{}{%
    directory,
  },
  [root, s sep=0.0mm %
    [egs2, s sep=0.0mm
      [corpus-name, s sep=0.0mm
        [enh1, s sep=0.0mm
          [conf, label=right:(configuration files)
          ]
          [local, label=right:(corpus-dependent)
          ]
          [data, label=right:(pre-processed Kaldi-style data)
          ]
          [dump, label=right:(speech features)
          ]
          [exp, label=right:(experiments)
          ]
          [enh.sh, file, label=right:(main script)
          ]
        ]
      ]
    ]
    [espnet2, label=right:(task and model definition)
    ]
    [tools, label=right:(tool installation)
    ]
  ]
\end{forest}
\end{minipage}
\caption{Directory structure of ESPnet-SE.}
\label{fig:se_dir}
\end{figure}
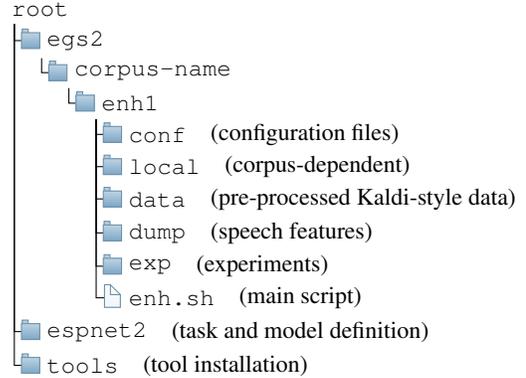

\begin{table*}
  \caption{Performance (PSEQ\,/\,STOI\,/\,SDR) on the CHiME-4 1-channel and 6-channel tracks.}
  \label{tab:chime4}
  \centering
  \begin{tabular}{l | c | c c c | c c c}
    \toprule
    \multirow{2}{*}{\textbf{Model}} & \multirow{2}{*}{\textbf{Track}} & \multicolumn{3}{c|}{\textbf{Dev (Simu)}} & \multicolumn{3}{c}{\textbf{Test (Simu)}} \\
     & & \textbf{PESQ} & \textbf{STOI} & \textbf{SDR (dB)} & \textbf{PESQ} & \textbf{STOI} & \textbf{SDR (dB)} \\
    \midrule
    Noisy Input (CH5) & \multirow{3}{*}{1-ch} & 2.17 & 0.86 & 5.78 & 2.18 & 0.87 & 7.54 \\
    T-F masking & & 2.55 & 0.91 & 10.36 & 2.46 & 0.89 & 11.61 \\
    Conv-TasNet & & 2.46 & 0.90 & 11.02 & 2.40 & 0.88 & 12.16 \\
    \midrule
    BeamformIt$^{\text{7}}$ & \multirow{2}{*}{6-ch} & 2.31 & 0.88 & 5.51 & 2.20 & 0.86 & 6.25 \\
    BLSTM MVDR & & \textbf{2.68} & \textbf{0.95} & \textbf{13.40} & \textbf{2.68} & \textbf{0.95} & \textbf{14.10} \\
    \bottomrule
  \end{tabular}%
\end{table*}
\subsection{Recipes}
\label{sec:recipes}
We provide various all-in-one recipes for several SE benchmark corpora, including WSJ0-2mix \cite{Deep-Hershey2016,Single-Isik2016} and its spatialized version \cite{Multi-Wang2018}, CHiME-4 \cite{CHiME4-Vincent2017}, REVERB \cite{REVERB-Kinoshita2016}, DIRHA \cite{DIRHA-Cristoforetti2014}, SMS-WSJ \cite{SmsWsj-Drude2019}, LibriCSS \cite{chen2020continuous}, LibriMix \cite{Librimix-Cosentino2020}, WHAM \cite{WHAM-Wichern2019} and WHAMR \cite{WHAMR-Maciejewski2019}.
Note that all these datasets are based on the WSJ \cite{wsj0} or Librispeech \cite{Librispeech-Panayotov2015} corpora, so we can use strong ESPnet-ASR pre-trained models for evaluation and joint optimization.
The directory structure of ESPnet-SE recipes is depicted in Fig.~\ref{fig:se_dir}.

All recipes follow a unified pipeline with a stage-by-stage processing style.
The function of each stage in \texttt{enh.sh} is described below:
\paragraph*{Stage 1:}
Corpus-dependent data is prepared in the \texttt{data} directory, following the Kaldi style.
\paragraph*{Stage 2:}
(Optional) Speed perturbation is performed on the training dataset\footnote{We found this is helpful for training ESPnet-SE models on data without reverberation, as shown in Table~\ref{tab:wsj0_2mix}.}.
\paragraph*{Stage 3:}
The pipe-style audio data in \texttt{wav.scp}, such as ``\texttt{sph2pipe -f wav some/wav/path |}'', is dumped to real audio files under the \texttt{dump} directory. The sampling rate and and audio format can also be changed if specified.
\paragraph*{Stage 4:}
Underlength and overlength samples are removed from both training and development sets.
\paragraph*{Stage 5:}
The input information of both training and development sets, including the sample length and number of channels, is collected for model training.
\paragraph*{Stage 6:}
Model training is performed.
\paragraph*{Stage 7 \& 8:}
Speech enhancement/separation is performed on the evaluation sets  to generate enhanced audios, and scores are calculated with different metrics, such as PESQ, STOI and SDR.
\paragraph*{Stage 9 \& 10:}
Decoding with a pretrained ASR model is conducted and scores are calculated.
\paragraph*{Stage 11 \& 12:}
(Optional) The trained model is packed and then uploaded to the Zenodo community\footnote{\url{https://zenodo.org/communities/espnet/}} via a specialized API \texttt{espnet\_model\_zoo}\footnote{\url{https://github.com/espnet/espnet_model_zoo}}. This makes it easy to share and reuse state-of-the-art models obtained with ESPnet.
\newline

For the joint optimization of SE and ASR models, Stage 6 above covers the training of both networks as a whole system. As a result, the decoding is directly performed with the trained model itself. 

\section{Experiments Evaluation}
\label{sec:exp}
To demonstrate the effectiveness of the proposed ESPnet-SE framework, we conducted several experiments with different datasets from speech enhancement to separation. In this section, we mainly introduce some experimental evaluations on the CHiME-4 speech enhancement challenge, single-channel WSJ0-2mix speech separation task and the spatialized version with multi-channel speech separation. Besides, the speech recognition experiments have also been conducted to show the functionality of our proposed ASR integration.

\subsection{Speech enhancement on CHiME-4}
We first conduct experiments on CHiME-4 \cite{CHiME4-Vincent2017}, a speech enhancement task where only one speaker exists in each sample.
The CHiME-4 dataset contains both simulated data and real recordings, based on the Wall Street Journal (WSJ0) corpus. The sampling rate of CHiME-4 data is 16 kHz. In our experiments, we only use the simulated data for training and evaluation, because there are no clean speech labels for the real recordings. The simulated dataset consists of a 15.1-hour training set, a 2.9-hour development set, and a 2.3-hour test set of 6-channel noisy speech. The 1-channel track of CHiME-4 only selects one of the six channels as input for evaluation, while the 6-channel track uses all channels as input.

We evaluate our implementation of the T-F masking model and TasNet on the 1-channel track and the neural beamformer (denoted as BLSTM MVDR) on the 6-channel track. We also evaluate BeamformIt \cite{Acoustic-Xavier2007} as a baseline for the 6-channel track. For both tracks, we take the 5-th channel (CH5) clean signal as the reference for computing all metrics, including PESQ, STOI and SDR. The results are shown in Table~\ref{tab:chime4}.
It shows that our ESPnet-SE models work well on both 1-channel and 6-channel tracks, with significant improvement over the baselines on all metrics.

Furthermore, we also evaluate the speech recognition performance of the ESPnet-SE models with an ASR model, which was pretrained on CHiME-4. The WER results on both simulation and real data are present in Table~\ref{tab:chime4_wer}. It can be observed that a simple combination of the neural beamformer and the pretrained ASR model can bring a significant performance improvement, compared to a single ASR model on the 1-channel track. And it also achieves better performance than the baseline model (6-channel track) in~\cite{CHiME4-Vincent2017}.
\begin{table}[t]
  \caption{WER (\%) on the CHiME-4 simulation and real data. All results were obtained with the same pretrained ASR model.}
  \label{tab:chime4_wer}
  \centering
  \begin{tabular}{ll | c c | c c}
    \toprule
    \multirow{2}{*}{\textbf{Track}} & \multirow{2}{*}{\textbf{Model}} & \multicolumn{2}{c|}{\textbf{Dev}} & \multicolumn{2}{c}{\textbf{Test}} \\
    & & \textbf{real} & \textbf{simu} & \textbf{real} & \textbf{simu} \\
    \midrule
    \multirow{3}{*}{1-ch} & Baseline~\cite{CHiME4-Vincent2017} & 11.6 & 13.0 & 23.7 & 20.8 \\
    & ESPnet-ASR & 10.9 & 12.6 & 19.5 & 19.9 \\
    & $\ \ $+ Conv-TasNet & 15.0 & 16.4 & 30.1 & 26.8 \\
    \hline
    \multirow{4}{*}{6-ch} & Baseline~\cite{CHiME4-Vincent2017} & 5.8 & 6.8 & 11.5 & 10.9 \\
    & ESPnet-ASR & 10.9 & 12.6 & 19.5 & 19.9 \\
    & $\ \ $+ BeamformIt\protect\footnotemark & 7.3 & 8.4 & 13.2 & 13.9 \\
    & $\ \ $+ BLSTM MVDR & \textbf{5.9} & \textbf{5.3} & \textbf{9.8} & \textbf{8.0} \\
    \bottomrule
  \end{tabular}%
\end{table}

\footnotetext{We use the BeamformIt toolkit implemented in \url{https://github.com/xanguera/BeamformIt} for evaluation.}

\begin{table*}
  \caption{Performance (PSEQ\,/\,STOI\,/\,SDRi\,/\,WER) on the WSJ0-2mix dataset. WERs are evaluated with the same ASR model.}
  \label{tab:wsj0_2mix}
  \centering
  \begin{tabular}{l | c c c c | c c c c}
    \toprule
    \multirow{2}{*}{\textbf{Model}} & \multicolumn{4}{c|}{\textbf{Dev}} & \multicolumn{4}{c}{\textbf{Eval}} \\
    & \textbf{PESQ} & \textbf{STOI} & \textbf{SDRi (dB)}  & \textbf{WER (\%)} & \textbf{PESQ} & \textbf{STOI} & \textbf{SDRi (dB)} & \textbf{WER (\%)}\\
    \midrule
    Mixture & 2.00 & 0.72 & - & - & 2.01 & 0.74 & - & - \\
    uPIT-PSM-relu \cite{Kolbaek2017Multitalker} & - & - & 9.4 & - & - & - & 9.4 & - \\
     uPIT-PSM-relu  & 2.96 & 0.89 & 10.3 & 31.9 & 2.92 & 0.90 & 10.2  & 29.1 \\
    \quad{+ speed perturb}  & 3.04 & 0.90 & 10.6 & 29.2 & 2.99 & 0.90 & 10.6  & 26.0 \\
    Conv-TasNet \cite{Conv-Luo2019} & - & - & - & - & - & - & 15.6 & - \\
    Conv-TasNet & 3.38 & 0.95 & 17.1 & 17.7 & 3.30 & 0.95 & 16.0 & 19.1 \\
    \quad{+ speed perturb} & 3.46 & 0.95 & 17.9 & 16.1 & 3.39 & 0.96 & 17.2 & 16.7 \\
    DPRNN \cite{luo2020dual} & - & - & - &  & - & - & 19.0  & - \\
    DPRNN & 3.43 & 0.95 & 17.8 & 17.2 & 3.42 & 0.97 & 17.9 & 16.1 \\

    \bottomrule
  \end{tabular}%
\end{table*}

\begin{table*}
  \caption{Performance (PSEQ\,/\,STOI\,/\,SDR\,/\,WER) on the spatialized WSJ0-2mix dataset.}
  \label{tab:spatialized_wsj0_2mix}
  \centering
  \begin{tabular}{l | c | c c c | c c c c}
    \toprule
    \multirow{2}{*}{\textbf{Model}} & \multirow{2}{*}{\textbf{Condition}} & \multicolumn{3}{c|}{\textbf{Dev}} & \multicolumn{4}{c}{\textbf{Eval}} \\
    & & \textbf{PESQ} & \textbf{STOI} & \textbf{SDR (dB)} & \textbf{PESQ} & \textbf{STOI} & \textbf{SDR (dB)} & \textbf{WER (\%)} \\
    \midrule
    Original Mixture & \multirow{4}{*}{anechoic} & 2.00 & 0.72 & 0.16 & 2.01 & 0.74 & 0.15 & - \\
    2-ch Deep clustering \cite{Multi-Wang2018} & & - & - & - & - & - & 12.9 & - \\
    IRM\,/\,IBM \cite{Multi-Wang2018} & & - & - & - & - & - & 12.7\,/\,13.5 & - \\
    2-ch BLSTM MVDR & & \textbf{3.48} & \textbf{0.97} & \textbf{20.6} & \textbf{3.43} & \textbf{0.97} & \textbf{20.5} & 34.4 \\
    \midrule
    Original Mixture & \multirow{6}{*}{reverberant} & 1.83 & 0.64 & -0.29 & 1.81 & 0.66 & -0.30 & - \\
    8-ch Oracle MCWF \cite{Multi-Wang2018} & & - & - & - & - & - & 10.9 & - \\
    IRM\,/\,IBM \cite{Multi-Wang2018} & & - & - & - & - & - & 11.9\,/\,12.7 & - \\
    8-ch BLSTM MVDR & & 2.45 & 0.81 & 10.19 & 2.40 & 0.82 & 10.05 & 70.3 \\
    $\ \ $+ Nara-WPE \cite{Nara-Drude2018} & & \textbf{2.67} & \textbf{0.83} & \textbf{11.55} & \textbf{2.67} & \textbf{0.84} & \textbf{11.94} & 56.7 \\
    $\ \ $+ DNN-WPE$^{\text{10}}$ & & 2.17 & 0.75 & 4.90 & 2.18 & 0.78 & 5.33 & - \\
    \bottomrule
  \end{tabular}%
\vspace{-0.3cm}
\end{table*}

\subsection{Speech separation and recognition on WSJ0-2mix}\label{subsec:wsj0-2mix}
WSJ0-2mix~\cite{Deep-Hershey2016} is the current benchmark dataset for the single-channel speech separation task. This dataset consists of a 30-hour training set, a 10-hour validation set, and a 5-hour test set of single-channel two-speaker mixtures without noise and reverberation. In our experiments, we use the WSJ0-2mix data with 8-kHz sampling rate for training and evaluation. With this benchmark, plenty of speech separation methods have been proposed in recent years, including deep clustering~\cite{Deep-Hershey2016}, PIT~\cite{Permutation-Yu2017}, TasNet~\cite{Tasnet-Luo2018,Conv-Luo2019}, and DPRNN~\cite{luo2020dual}.

Among these methods, we compare our implementation of three representative methods in the time-frequency domain (uPIT~\cite{Kolbaek2017Multitalker}) and time domain (Conv-TasNet~\cite{Conv-Luo2019} and DPRNN) to those from prior studies. The results are shown in Table~\ref{tab:wsj0_2mix}.
As we can see, our implementations achieve promising results on the WSJ0-2mix, with a performance comparable to originally reported.

Interestingly, we also find that speed perturbation \cite{Audio-Ko2015}, a broadly used method for data augmentation in ASR, is also helpful in training the ESPnet-SE model. By applying speed perturbation with speed factors of 0.9, 1.0 and 1.1, we can observe an absolute SDR improvement of $1.2$ dB on the TasNet model and 0.4 dB on the T-F masking model.

In addition, we also evaluate these well-trained speech separation models with an ESPnet-ASR model pretrained on WSJ.
The WER of the pretrained ASR model on WSJ Dev93 and Eval92 are 9.9\% and 7.0\% respectively.
By simply combining the ESPnet-SE model and ASR model, we obtain a surprisingly good performance on the WSJ0-2mix dataset.
As shown in the last column in Table~\ref{tab:wsj0_2mix}, the combination of Conv-TasNet and ASR models achieves 16.7\% WER on the evaluation set, and the DPRNN achieves better performance with 16.1\% WER.

\subsection{Speech separation on spatialized WSJ0-2mix}
The spatialized WSJ0-2mix \cite{Multi-Wang2018} is a multi-channel speech separation dataset, extended from the original WSJ0-2mix dataset.
A room impulse response (RIR) generator\footnote{\url{https://github.com/ehabets/RIR-Generator}} based on the image method \cite{Image-Allen1979} is used to spatialize the WSJ0-2mix dataset.
It contains an anechoic and a reverberant version, both of which consists of 8-channel two-speaker mixtures.
In our experiments, we use the spatialized WSJ0-2mix data with 8-kHz sampling rate for training and evaluation.

We evaluate our implementation of the neural beamformer (denoted as BLSTM MVDR) on both anechoic and reverberant conditions. For evaluation on both anechoic and reverberant data, we use the anechoic signal of each speaker at the first channel as the reference, and the results are shown in Table~\ref{tab:spatialized_wsj0_2mix}.
We can observe that in the anechoic condition, our implemented neural beamformer outperforms the prior work \cite{Multi-Wang2018}, with more than 7dB SDR improvement.
In the reverberant condition\footnote{Note that the results in the reverberant condition is not fully comparable, because a different reference signal is used in \cite{Multi-Wang2018}.}, our model still achieves a good performance on a par with the prior work \cite{Multi-Wang2018}.
And with an additional WPE module for dereverberation, the performance can be further improved.

Finally, we also evaluate the performance with ASR integration.
The same ASR model trained on WSJ is used for ASR evaluation.
The results are present in the last column in Table~\ref{tab:spatialized_wsj0_2mix}.

\begin{figure*}
\centerline{\includegraphics[width=\textwidth]{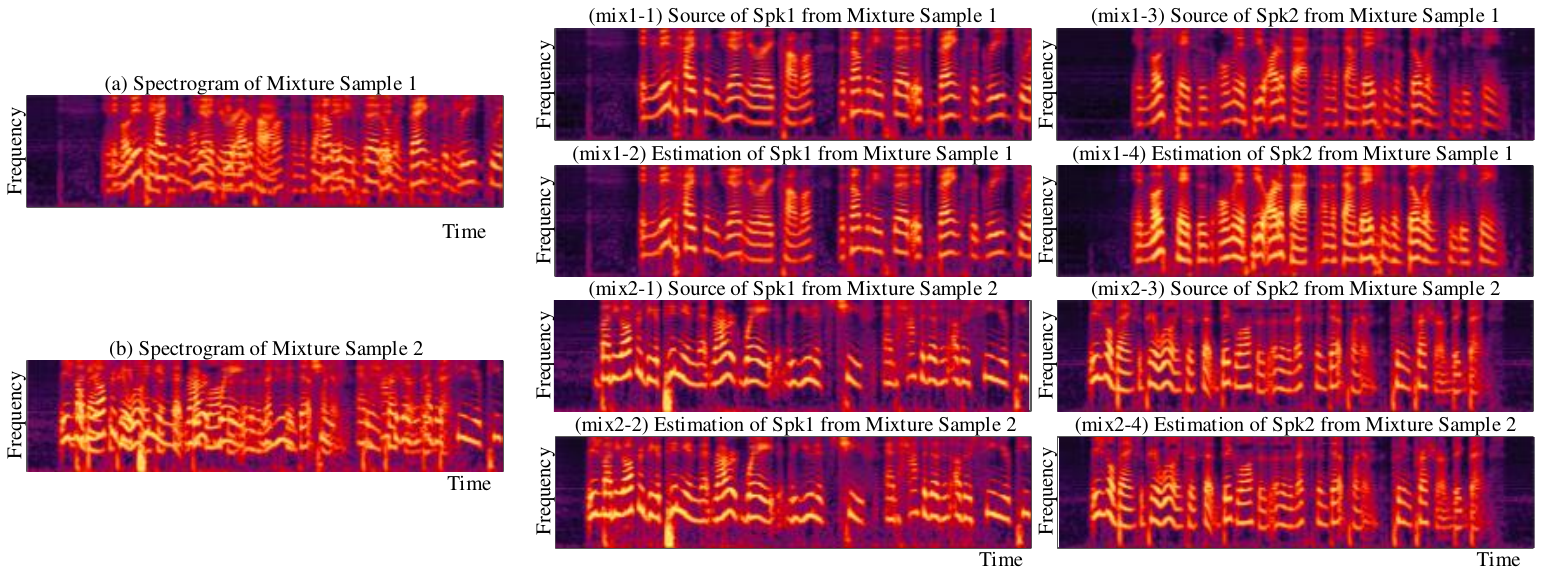}}
\caption{Visualization of two same-gender mixtures from the WSJ0-2mix test set. Mixture 1 consists of two males and the average SDR of this sample is 18.54 dB, while Mixture2 consists of two females, with the average SDR of 12.88 dB.}
\label{fig:spectrograms}
\end{figure*}
\subsection{Joint training of SE and ASR}
\begin{table}[t]
\centering
\caption{The performance on the WSJ0-2mix datasets with joint training of E2E-SE and ASR module.}
\label{tab:joint}
\begin{tabular}{c|ccc}
\toprule
\textbf{Model} & \multicolumn{3}{c}{\textbf{Eval}} \\
(SE + ASR) & \textbf{SDRi} & \textbf{CER} & \textbf{WER} \\\midrule
TF-masking + Transformer & 8.5 & 18.6 & 33.8\\
Conv-TasNet + Transformer & 14.7 & 8.1 & 15.8\\
\bottomrule
\end{tabular}%
\end{table}

\footnotetext[10]{The DNN-WPE module is still under development.}
As illustrated in Fig.~\ref{fig:integration_with_asr}, our entire system could be optimized with the final ASR loss, while the SE loss  could also be an additive option to train the same architecture. Here,
we showed two example experiments on the WSJ0-2mix dataset trained with both SE (speech separation here) and ASR loss. The results are shown in Table~\ref{tab:joint}. The model we use is the same as each module of T-F masking or Conv-TasNet separation network and ASR model used in section~\ref{subsec:wsj0-2mix}, while it is directly trained with WSJ0-2mix rather than WSJ from scratch. 

From the results we could observe that, the joint optimization works well for the downstream application. With the training only on the WSJ0-2mix, comparable or better WER has been obtained compared with the strong single-speaker ASR trained on the whole WSJ dataset (see in Table~\ref{tab:wsj0_2mix}). However, the separation performance gets worse than the isolated separation model. This tendency is also observed from some other works~\cite{Speech-Subramanian2019,von2020multi}.

In addition, we are working on the end-to-end joint training with ASR loss only. 
Previous studies~\cite{MIMO-Chang2019,End-Chang2020,subramanian2019investigation} have shown promising results in this direction, and all of them got the implementation with ESPnet.
Some ongoing work by us is to reproduce these works. It is optimistic to make it work since the ESPnet-SE shares the similar design ideas and the bases with ESPnet.

\subsection{Visualization of enhanced spectrogram}
In this section, we visualize the spectrograms of both input and enhanced signals to demonstrate the power of our ESPnet-SE model, which can be easily generated within the ESPnet-SE recipe. Here we adopt the well-trained DPRNN model for speech separation.
We randomly selected two same-gender speech mixtures from the WSJ0-2mix test set. The spectrograms of the mixed signals, the corresponding clean source signals, and the estimated source signals are shown in Fig.~\ref{fig:spectrograms}.
As we can see, our implemented DPRNN model can generate very good estimates of the source signals from the mixture.

\section{Conclusions}
\label{sec:conclude}
In this paper, we present an end-to-end speech enhancement and separation toolkit, named ESPnet-SE, which can be tightly integrated with ASR.
The toolkit is an extension of the open-source toolkit ESPnet, and supports various state-of-the-art speech enhancement and separation models with rich features.
It also provides plenty of all-in-one recipes with a unified stage-by-stage processing pipeline, including data preparation, feature extraction, model training and evaluation. A wide range of benchmark datasets are covered in the current toolkit.
A core feature that differentiates this toolkit from others is the capability of ASR integration. With the unified design in ESPnet-SE, we can easily integrate speech enhancement with ASR, such as model evaluation with the ASR criterion and joint optimization with ASR.

In the future, we will support more speech enhancement and separation models, such as MIMO-Speech \cite{MIMO-Chang2019}, which is tightly integrated with ASR. We will also support target-speaker speech extraction models like speaker beam \cite{SBeam-Delcroix2019} in the future. In addition, we will work on a more flexible design to enable the integration of external E2E-SE models with ESPnet-ASR.

\section{Acknowledgement}
A part of this work was studied during JSALT 2020 at JHU, with support from Microsoft, Amazon, and Google. 

\bibliographystyle{IEEEbib}
\bibliography{refs}

\end{document}